
\magnification=1200
\baselineskip=20pt
\centerline{\bf COMMENT ON `` ANOMALY-INDUCED MAGNETIC SCREENING}
\centerline{\bf IN (2+1)
 DIMENSIONAL QED AT FINITE DENSITY"}

\vskip 1cm

\centerline{by}

\vskip 1cm

\centerline{C. R. Hagen}
\centerline{Department of Physics and Astronomy}
\centerline{University of Rochester}
\centerline{Rochester, NY 14627}

\vfill\eject

In a recent work$^1$ it has been asserted that a Meissner effect can be
derived from a conventional QED in 2 + 1 dimensions provided that a chemical
potential term is included.  In this note it will be pointed out that there
are two specific (and independent) criticisms which invalidate that result.

The first point has to do with the claim [Eq. (23)] that the
``Maxwell equation"
$$J_i = {1 \over e^2}\ \epsilon^{ij} \partial_j B(x)
\eqno(1)$$
can be freely inserted into the alleged curl equation [Eq. (22)]
$$\langle \epsilon^{ij} \partial_i J_j (x) \rangle =
 - | \mu | B(x) \left[ {1 \over 2 \pi} +
O \left( B/\mu^2 \right) \right] \eqno(2)$$
(where $\mu$ is the chemical potential) to obtain immediately a
Yukawa-type equation for $B(x)$.  In fact Eq. (1) is not valid since the
vector potential $A_\mu (x)$ and the associated magnetic field $B$  are
{\it external} fields while $J_i (x)$ is a current operator which is
bilinear in the quantized spinor fields.  Thus, for example, the two sides
of Eq. (1) have different commutation relations with the underlying field
operators.  Although ref. 1 {\it does} start with a Lagrangian which seems to
imply a quantized electromagnetic field, Eq. (2) clearly equates $B (x)$ to
a $c$-number.  Furthermore, the derivation of (2) presented in ref. 1 can
only be carried out for classical (i.e., external) electromagnetic fields.
It must also be noted that (1) is a Maxwell-type equation only if a term
proportional to the time derivative of the electric field is neglected, a
step which cannot be done, of course, for a quantized electromagnetic
field.

The second point has to do with the derivation of (2).  The approach of
ref. 1 seems needlessly involved (and difficult to follow) particularly
since only minor modifications have to be made to ordinary covariant
perturbation theory in order to calculate the $\ell h s$ of Eq. (2).  To
demonstrate this one writes
$$\langle J^\mu (x) \rangle = \int \Pi^{\mu \nu} (x - x^\prime ) A_\nu
(x^\prime) dx^\prime$$
where $\Pi^{\mu \nu} (x)$ is the current correlation function.  To lowest
order in the coupling its Fourier transform can be written in terms of the
fermion propagator $S (p)$ as$^2$
$$\Pi^{\mu \nu} (q) = i \int {dp \over (2 \pi )^3} \
{\rm Tr} \left[ \gamma^\mu \left\{
S(p+q/2)\gamma^\nu S (p-q/2) +
{\partial \over \partial p_\nu}\ S (p) \right\} \right] \eqno(3)$$
which by current conservation has the form
$$\Pi^{\mu \nu} (q) = \left( g^{\mu \nu} - q^\mu
q^\nu/q^2 \right) \Pi (q) + \left(
\delta^\mu_0 q^\nu + \delta^\nu_0 q^\mu -
q^\mu q^\nu q_0/q^2 -
\delta^\mu_0 \delta^\nu_0 q^2/q_0 \right) \Pi^\prime (q) \quad .$$
(The function $\Pi^\prime (q)$ must be allowed, but vanishes when the
chemical potential goes to zero.)  This implies the result
$$\langle \epsilon^{ij} \partial_i J_j (x) \rangle =
\int \Pi (x-x^\prime) B(x^\prime) dx^\prime$$
which (for sufficiently slowly varying magnetic fields) becomes
$$\langle \epsilon^{ij} \partial_i J_j (x) \rangle = \Pi
(q = 0) B \quad .$$
Although the chemical potential modifies the form of the propagator$^3$
$S(p)$, it is nonetheless easy to see that because of the Ward identity the
two terms in (3) exactly cancel$^4$ for $q=0$ and the $rhs$ of (2)
necessarily vanishes (up to terms which depend on derivatives of the
external field).
 It is worth noting that the crucial second term on the
 $rhs$ of Eq. (3) has its origin in the careful gauge invariant definition
 of the current, an issue which is not discussed in ref. 1$^5$.
 One thus concludes that the calculation of the two
dimensional curl of the vacuum expectation value  of the current operator
as presented in ref. 1 is not correct.

\noindent {\bf Acknowledgment}

This work is supported in part by the U.S. Department of Energy Grant No.
DE-FG-02-91ER40685.
\vfill\eject

\noindent {\bf References}

\item{1.} S. Forte, Phys. Rev. Lett. {\bf 71}, 1303 (1993).

\item{2.} K. Johnson in
 Brandeis Summer Institute in Theoretical Physics, 1964 v. 2,
Prentice-Hall, Inc., Englewood Cliffs, NJ.

\item{3.} E. V. Shuryak, Phys. Reports {\bf 61}, 71 (1980).

\item{4.} This argument would fail, however, in the case of the Schwinger
model since the expression (3) is infrared divergent at $q=0$
 in that case.  Because of this complication  it is
necessary to give the fermion a mass which subsequently is allowed to go to
zero.  The cancellation between the two terms of (3) thus would not occur
because the limits $q \rightarrow 0$ and the fermion mass going to zero
must be taken in opposite order for these two terms.  Since there is an
additional spatial dimension in the case at hand, there is no corresponding
infrared problem and the conclusion concerning the vanishing of $\Pi (q=0)$
is correct.
\item{5.} The lack of gauge invariance is apparent in its detailed dependence
on the gauge dependent eigenvalues $\lambda_k$ of the Dirac operator.  The
remark made in the first work cited in footnote 9 of ref. 1 concerning
gauge invariance in the $\zeta$-function regularization method is also quite
relevant.  Note in particular Eq. (5.17), a result which clearly
establishes the absence of gauge invariance in the calculation of ref. 1.

\end